\documentclass{phb-proc4-auth}

\usepackage{graphicx}
\usepackage{amssymb}

\begin{document}
\begin{frontmatter}


\journal{SCES '04}


\title{Do quantum dots allow one access to pseudogap Kondo physics?}

%
%
%
%
%
%

\author{{\corauthref{1}}John Hopkinson}
\author{\hskip-0.21pc, Karyn Le Hur,}
\author{and {\'{E}}milie Dupont}

%
 
\address{D{\'e}partement de Physique, Universit{\'e} de Sherbrooke, Sherbrooke, Qu{\'e}bec, Canada}

%
%
%
%

\thanks{This work was supported by NSERC, FQRNT and CIAR.}

%
%
%
%

\corauth[1]{Corresponding Author: Universit{\'e} de Sherbrooke, QC, CA Phone: (819) 821-8000 ext. 3043 Fax: (819) 821-8046}


\begin{abstract}
For the last decade{\cite{revival}}, tunable quantum dot systems have allowed the investigation of Kondo physics wherein the quenching of a single spin on an artificial atom affects the conductance.  The pseudogap Kondo model (pKm) featuring a density of states $\rho(\epsilon)$ = C$|\epsilon|^{r}$, introduced by Withoff and Fradkin{\cite{withoff}} in 1990 was predicted to exhibit Kondo-like physics above a critical value of the Kondo coupling, J$_c$, which several groups have shown by numerical renormalization group (RG) is finite for r$< \frac{1}{2}$.  Gonzalez-Buxton {\it{et al}} {\cite{gonzalez}} showed that the strong coupling limit of the particle-hole symmetric model leads to a non-trivial $\frac{\pi (1-r)}{2}$ phase shift at low temperatures indicating incomplete screening of the local moment, while away from particle-hole symmetry (p-hs) one generically flows towards a ground state with $\delta \sim \pi$.  We examine the implications of this model for quantum dots whose leads are Fermi-liquid-like, yet possess a tunneling density of states (TDOS) which is suppressed at the Fermi energy as a power law.  
\end{abstract}

%
%



\end{frontmatter}

%
%
%
%

Many of the unusual features arising in strongly correlated electron systems have been attributed to the proximity to quantum critical points.  In particular, heavy fermion systems are thought to feature a competition between RKKY and Kondo interactions, within the context of a Kondo lattice model, leading to (an increasingly heavy) metallic antiferromagnetic phase which is suppressed as a function of pressure, chemical substitution or magnetic field to produce a cone of non-Fermi liquid physics preceeding a (decreasingly heavy) metallic state.  Unfortunately, despite decades of intensive research on the Kondo lattice model, the detailed nature of its solution remains a subject of some debate.  A simpler model exhibiting a quantum phase transition between a state with local moments and spin-screened state, the pKm, was introduced in Ref. 2 and studied extensively by subsequent groups{\cite{gonzalez,chen,ingersent1}}.  This paper explores the possibility that this model could be experimentally realizable in a quantum dot system.



{\it{Resonant level limit}} We consider the small dot, weak tunneling limit in which the dot is a collection of discrete levels of average spacing $\delta$E, an order of magnitude smaller than the charging energy, E$_c$ (${\mathcal{O}}$($\frac{e^2}{2C}$)), of the dot (e: electron charge; C: capacitance of dot-lead system).  The energy to add or subtract an electron from the dot can be estimated as $\delta$E$^{\pm}_c$ = E$_c$(Q$_G$,n+1) - E$_c$(Q$_G$, n)  = (n+$\frac{1}{2}$ - C$_G$V$_G$)$\frac{e^2}{C}$, where n is the initial(final) number of electrons on the dot for +(-), V$_G$ is an external gate voltage coupled capacitively (C$_G$) to the dot.  A resonant condition periodic in the number of electrons on the dot corresponds to $\delta$E$_c^{+(-)}$ = 0, $\delta$E$_c^{-(+)}$ = $\frac{e^2}{C}$.  For temperatures satisfying $\delta$E$<$T$<$E$_c$, the physics of the dot is dominated by Coulomb blockade.  For temperatures below $\delta$E, one can write the current through the dot in the single level Landauer-B{\"u}ttiker (elastic scattering) form,
\begin{equation}
I = \frac{2e}{h}\int_{-D}^{D} \frac{4(f_L(\epsilon)-f_R(\epsilon))\Gamma_L\Gamma_R|\frac{\epsilon}{D}|^{2r}d\epsilon}{(\epsilon-\epsilon_d)^2 + (\Gamma_L + \Gamma_R)^2|\frac{\epsilon}{D}|^{2r}}.
\end{equation} 
$f_{L(R)}(\epsilon)$ is the Fermi function of the left(right) lead, we have assumed that we are at sufficiently low T that the intervals $\{\pm D,\pm\delta E\}$ contribute weakly so we can neglect multiple states and have chosen a power-law dot-lead hybridization $ \Gamma(\epsilon) = \Gamma_0|\frac{\epsilon}{D}|^r$ where $\Gamma_L = \Gamma_R = \Gamma_0 = \pi(r+1)t_kt_{k'}/(2D)$ ($t_k$: dot-lead hopping, $|\epsilon|<D$). Implicit is the assumption of equivalent leads and dot-lead couplings, and that the effect of electron-electron interactions within the leads is completely captured by a renormalization of the TDOS of the conduction electrons, (in the spirit of the one-dimensional (1 D) wire RG of Nazarov and Glazman{\cite{nazarov}}).  Using (1) we can calculate the bare conductance (G$_0$ = $\frac{dI}{dV}|_{V\rightarrow 0}$). Notice that at T=0 the manifestation of Coulomb blockade is an infinitely sharp transmission amplitude at resonance ($\epsilon_d = 0$).

{\it{Kondo limit}} For T$<\delta E$, directly between two resonant points, with an odd number of spins on the dot, the symmetric condition $\delta$E$_c^{(+)}$= $\delta$E$_c^{(-)}$ = $\frac{e^2}{2C}$ allows one to write $H = \sum_{k\alpha\sigma}(\epsilon_k c_{k\alpha\sigma}^{\dagger} c_{k\alpha\sigma} + (t_{k\alpha}c_{k\alpha\sigma}^{\dagger}d_{\sigma} + h.c.)) + \epsilon_dn_d+ Un_{d\downarrow}n_{d\uparrow}$,
where c$_{k\alpha\sigma}$ destroys a conduction electron of momentum k, spin $\sigma$, from lead $\alpha$ and we have defined an effective on-site Hubbard repulsion U = -2$\epsilon_d$, where $\epsilon_d$ is the energy of the highest occupied level on the dot.  This model is the pseudogap Anderson model (pAm) studied in Ref. 3 with a reduced r{\'e}gime of validity (D$\rightarrow$E$_c$ in the energy summation).  In the limit $-\epsilon_d$ (n odd dot energy), U + $\epsilon_d$ (energy of 1st excitation) $>>$ $\Gamma_0$, T the Schrieffer-W{\"o}lff transformation leads to{\cite{gonzalez}} $H_K = \sum_{k,k'}(J_{k,k'}c_{k\alpha}^{\dagger}\frac{\sigma_{\alpha\beta}}{2}c_{k'\beta}S + V_{k,k'}c_{k\sigma}^{\dagger}c_{k'\sigma})$,
where the Kondo coupling, J$_{k,k'}$ = 2($\frac{1}{|\epsilon_d|} + \frac{1}{|U+\epsilon_d|}$)t$_k$t$_{k'}$, and the potential scattering V$_{k,k'}$ = $\frac{1}{2} (\frac{1}{|\epsilon_d|}- \frac{1}{|U+\epsilon_d|}$)t$_k$t$_{k'}$ vanishes at the symmetric point U=-2$\epsilon_d$.

 Following Glazman and Ra\u{\i}kh{\cite{raikh}}, 
the effects of correlation in the leads (r=0 for free electrons) is to decrease the effective conductance relative to unitarity at the Kondo limit as G = G$_0$sin$^2$($\delta$), a result one might have anticipated from Kane and Fisher's{\cite{kanef}} repulsive interactions in Luttinger liquids (there G$_0$ $\rightarrow$ gG$_0$, g: Luttinger parameter).  In the Kondo limit, Ref. 3 found that at p-hs, the pKm flows to a strong coupling fixed point corresponding to the U=0 limit of the pAm provided $\Gamma$ exceeds a critical value which increases monotonically with r to diverge at $\frac{1}{2}$.  The phase shift of this model at the Fermi energy is $\delta = \frac{\pi}{2} (1-r$sgn$(\epsilon))${\cite{gonzalez,chen}}.  However, this strong coupling fixed point is unstable with respect to the potential scattering term V.  At the symmetric point, V=0, one should{\cite{ingersent1}} enter the particle-hole symmetric fixed point below the Kondo scale T$_{\chi} \sim \delta E(\frac{J-J_c}{J})^{\frac{1}{r}}$.   If V$\ne$0, for 0$<$r$<$r$^*$=0.375, the initial flow is towards the p-hs fixed point.  Below a temperature scale T$_{\chi'} = |\frac{V}{\delta E}|^{\frac{1}{r}}$T$_{\chi}$ the problem flows to an asymmetric strong coupling fixed point{\cite{ingersent1}} whose phase shift{\cite{gonzalez}} about the Fermi energy is $\pi$.  While such a phase shift would be opaque to conductance measurements, it is interesting to note that in Ahanorov-Bohm experiments{\cite{yang}} a flux corresponding to a phase-shift of $\pi$ has been apparently observed, and remains unexplained.

In Fig. 1 the effect of p-hs breaking is sketched.  While one expects to see evidence of Kondo physics (r$<$r$^*$) at low temperatures of height comparable to the resonant peak, the width of the peak (in V$_G$) will similarly sharpen at T$\rightarrow$0 to approach a $\delta$-function at T=0, unlike at r=0.  While single-walled metallic carbon nanotubes can be created with a sufficiently small power law TDOS, their 1 D nature introduces new complications. 
The relation of these results with those of 1 D treatments{\cite{1dstuff}} where to our knowledge no critical Kondo coupling has yet been found will be discussed elsewhere{\cite{us}}. However, a small power-law TDOS (due to interactions between electrons) could be obtained with 2 D leads with a non-negligible resistance{\cite{slot}}. 

%
%
%
%

\begin{figure}
\includegraphics[scale=0.45]{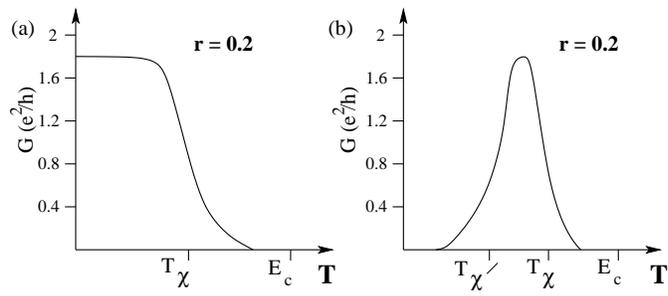}
\caption{\label{figure}A schematic sketch for r=0.2 of the conductance of the Coulomb blockade valley (a) at p-hs U=-2$\epsilon_d$; (b) for V$\ne$0.}
\end{figure}
%
%
%
%


\end{document}